\documentclass[aps,prb,twocolumn,showpacs]{revtex4}

\begin{document}

\title{
    About possible Phonon to Magnon alignment in 2 dimensions and 
    theory of superconductivity in Copper-Oxide planes.
}

\author{ Daniel~L.~Miller }
\address{
  Intel Electronics Ltd., 2200 Mission College, RNB-3-0, 
  Santa Clara, CA 95054-1549 
}

\email{Daniel.Miller@Intel.com}

\begin{abstract}
   We suggest that the phonon dispersion in cuprates becomes strongly anisotropic due to interaction 
   with spin waves; moreover the phonon dispersion becomes singular along $|k_x|=|k_y|$ directions.
   This would allow more electrons to form Cooper pairs and increase temperature of the superconducting
   transition. The interaction of phonons with spin waves is more important than the interaction of 
   phonons with free electrons, because spin waves do not have the Fermi surface constrain. 
\end{abstract}
\pacs{71.36.+c, 74.72.-h, 74.20.-z, 74.25.Kc 
\\ \hspace*{5.1in} to J.S.
}

\maketitle

The motivation for this work is to look for phonon-mediated mechanism of the high-temperature superconductivity 
which would give d-wave order parameter observed experimentally\cite{Shen-1993,Wollman-1993}, particularly by 
scanning tunneling imaging of non-magnetic dopants.\cite{dwave-Ni-2001} This letter will be focused on the dispersion 
of longitudinal acoustic phonons interacting with spin waves. This interaction has been introduced earlier to explain 
Raman spectra of high $T_c$ cuprates\cite{2Mag-Raman-1996}; it is similar to interaction of optical phonons with 
magnons\cite{Lorenzana-1995} seen possibly in midinfrared absorption of cuprates\cite{Perkins-1993,Struzhkin-2000}, 
Raman spectroscopy of Spin-Peierls systems \cite{2Mag-Raman-1997}, and reflectivity of CuO crystals\cite{spin-TO-2000}.  
We basically undertake a search for new anisotropic phonon modes seen in photoemission.\cite{Lanzara-Bogdanov-2001}

The electron-phonon interaction in the Fermi-liquid theory does not lead to interaction between phonons and 
spin waves. The phonon propagator is renormalized by the polarization bubble of electron gas, 
the latter is proportional to the electron density in the same way as conductivity and diffusion constant.
However electron preserves the spin when interacting with phonon so spin waves cannot interact with phonons.

Alternative description of spin waves - magnons can be derived from the Heisenberg Hamiltonian, just by 
attributing local spins to sites. In general the interaction energy $J\sim U^2/t$, where $U$ is the charging 
energy of local electron and $t$ is hopping matrix element. This matrix element depends exponentially on 
inter-atomic distance and so does the spin interaction $J$. We will show later that the spin-phonon interaction 
splits the dispersion of the longitudinal two-dimensional phonons 
and new branch becomes singular and anisotropic, namely it goes like
\begin{equation}
    \omega \sim ({1\over J} {dJ \over da})^2 {1 \over |\cos (k^xa) - \cos(k^y a)|}
\label{eq:intro.010}
\end{equation}
and the divergence is cut off by the condition $\omega \ll  J [\sin^2(k^xa/2)+ \sin^2(k^ya/2)]$, where
$a$ is the lattice constant of square lattice. The simple form Eq.~\ref{eq:intro.010} is valid for 
$\omega \gg \omega_0(\vec k)$, where $\omega_0(\vec k)$ is the dispersion law of longitudinal acoustic phonons.

The extra lattice stiffness comes out of zero-point energy of spin waves. The effect is similar to the nature
of attraction between bodies (Van der Waals forces) due to zero-point oscillations of electromagnetic 
waves.\cite{DLP-1961} By analogy, one can compute magnetostriction force due to polarization of
magnon vacuum. However this force would compete with all other strains in solid state and would be hardly 
observable. In the same time dynamical $\omega>0$ part of the magnon polarization operator has dramatic effect
on phonon dispersion. The right way to integrate out all almost standing 
magnon waves aligned with phonon wave is to solve the Dyson equation as it is done below by making use of 
zero temperature diagrams.

Two main ingredients are necessary to conduct the calculation: the spin-spin correlation function and spin-phonon
Hamiltonian (also known as magneto-elastic Hamiltonian). These two are not independent, for example
taking into account next-neighbor exchange interaction would affect both spin-spin correlation and spin-phonon
interaction. The subject is even more complicated in our ``target'' systems - high $T_c$ cuprates; their spin-spin 
correlation function depends on doping and changes across the antiferromagnet to superconductor 
transition\cite{Brinckman-Lee-2002}. Fortunately, the spin waves polarization operator contains $\omega$ integration
which completely removes this dependence, so our result Eq.~\ref{eq:intro.010} is robust
and should work even if the ground state does not have magnetic order at zero temperature.

According to the BCS theory\cite{three-authors-63} the energy gap $\Delta_{\vec k}$ is proportional to the 
phase volume available for scattering for two electrons with moments $\vec k$ and $-\vec k$ on the Fermi surface, 
when they loose energy $\omega$. The singular phonon dispersion leads to dramatic consequences for 
electron-electron pairing. The half-filled tight-binding model $E = t\cos(k^xa)+t\cos(k^ya)$ has rectangular 
Fermi surface $k^y = \pm \pi/a \pm k^x$. Assume that electron is on the line $k^x + k^y = \pi/a$. Then the
interaction with dispersion Eq.~(\ref{eq:intro.010}) will keep electron on the same bond 
$k^x_1 + k^y_1 = \pi/a$, since it allows the process with $|k^x_1 - k^x| = |k^y_1 - k^y|$. Twice more scattering out
phase volume is available from hot spots $(0,\pm \pi/a)$, $(\pm \pi/a, 0)$. For example from the point 
$\vec k = (0, \pi/a)$ the electron can be scattered out by new phonons to both 
$k^x_1 + k^y_1 = \pi/a$ and $ k^y_1 = k^x_1 - \pi/a$ bonds. The order parameter becomes anisotropic
$\Delta_{\vec k \in h.s.} = 2 \Delta_{\vec k \notin h.s.}$. Phonons Eq.~(\ref{eq:intro.010}) allow deviations from
the exact law $|k^x_1 - k^x| = |k^y_1 - k^y|$ and the Fermi surface is not rectangular for non-zero chemical 
potential so real life $\Delta_{\vec k }$ is smooth function having maxima at hot spots.

The alignment of phonons to the crystalline lattice is suggestive solution to the HTSC problem. It allows to interacting 
electrons to go up $\hbar\omega_{\text{max}}$ out of the Fermi surface ( below $\hbar=1$), increasing both the energy gap and 
transition temperature. Here $\omega_{\text{max}}$ is maximal (cutoff) frequency in Eq.~(\ref{eq:intro.010}); 
it sets up new energy scale between phonons and magnons; $\omega_{\text{max}}$ is much larger than the 
Debye frequency $\omega_D$ and much lower than the spin exchange energy $J$. The accurate 
calculation of $\Delta_{\vec k}$ requires regularization of the singularity in Eq.~(\ref{eq:intro.010}) and it is not
yet done.

Assume that the antiferromagnetic coupling constant $J>0$ depends on distance
\begin{equation}
   {\cal H} = \sum_{\langle ii' \rangle } J\vec S_i \vec S_{i'} = \sum_{\langle ii' \rangle } 
   (J + { d J\over d a} (\vec u_i - \vec u_{i'})\vec a_{i'\,i} )\vec S_i \vec S_{i'}
\label{eq:letter.010}
\end{equation}
where each pair of adjacent sites $\langle ii' \rangle$ on square lattice is counted once, 
vector $\vec a_{i'\,i}$ points from $i'$ to $i$,
$\vec u_i$ is displacement of $i$th site, and $\vec S_i$ is the spin on $i$th site.

Consider chain of sites counted along $x$ axis for given $y$, then the interaction term is
\begin{equation}
   \sum_{i}  (u^x_i - u^x_{i-1}) \vec S_i \vec S_{i-1}
   = \sum_{i}  u^x_i  \vec S_i ( \vec S_{i+1} - \vec S_{i-1} )\;.
\label{eq:letter.020}
\end{equation}
For the first glance this interaction vanishes because it is proportional to 
$\vec S \nabla_x \vec S = \nabla_x \vec S^2 = 0$. Extracting the following identity from the spin product
\[
    [\vec S_{i+1}+\vec S_{i-1}][\vec S_{i+1}-\vec S_{i-1}] = 0
\]
we will find interaction being proportional to higher power of derivatives. After some algebra
\begin{equation}
    {\cal H} =   - {1\over 2} {dJ \over da} \sum_{\langle ii' \rangle }
   (\vec u_i - \vec u_{i'})\vec a_{i'\,i} (\vec S_{i'} - \vec S_i)^2
\label{eq:letter.030}
\end{equation}
that can be verified by expansion of the square term. The magneto-elastic energy derived 
here is invariant under spin rotation as opposite to the so-called relativistic term\cite{gurevich-melkov-1996}. 
The relativistic part of the magneto-elastic energy is responsible for acoustic Faradey's 
effect - rotation of phonon polarization around magnetization axis or Neel's vector, observed recently in two 
dimensions.\cite{Ganiev-Kuzavko-1992} It will not be discussed here because i) the effect does not exist in
disordered phase, ii) it is less important for large $\vec k$.

For each bond $j=\langle ii' \rangle $ we will introduce the time dependent phonon field 
$\phi(j)=(\vec u_i - \vec u_{i'})\vec a_{i'\,i} \sqrt{ \rho } $, where $\rho$ is the material density. 
For longitudinal phonons the field is quantized as
\begin{equation}
    \phi(\xi) = 
    {1\over {\cal N}^{1/2}} \sum_{\vec k} { 2\kappa^x_{\vec k} k^x/k \over \sqrt{2\omega_0(k)}}
     \{ b_{\vec k} e^{i\vec k \vec r_j - i\omega_0(k) t } 
     +  c.c.\}
\label{eq:letter.040}
\end{equation}
for bond $j$ parallel to x-axis, and $y$-components of $\vec k$ should be taken if the bond 
$j$ is parallel to $y$ axis,
besides $[b_{\vec k}, b^\dagger_{\vec k'}] = \delta_{\vec k \vec k'}$ are boson operators,  $\xi=(j,t)$,
 and
\begin{equation}
      \vec \kappa_{\vec k} = (\sin(k^x a/2), \sin(k^y a/2))\;.
\label{eq:letter.045}
\end{equation}
The propagator for this real field is 
\begin{equation}
   D(\xi,\xi') = - i \langle T \phi(\xi)\phi(\xi')\rangle\;,\quad 
   D^{(0)}_{\omega, \vec k} = 
  {  (\vec \kappa_{\vec k} )^2
    \over \omega^2 - \omega_0^2(\vec k) + i \delta}
\label{eq:letter.050}
\end{equation}
In further computations we will not need the explicit form of $\omega_0(\vec k)$, and the only assumption is that
$0\le \omega_0(\vec k)\le \omega_D$, where $\omega_D$ is the Debye frequency. In other words we need 
$\omega_0(\vec k)$ to be limited for all $\vec k$.

The quantization of spin waves is convenient to start with choice of $z$ axis parallel to the 
direction of the magnetic order parameter; then 
the commutation relation of spins $ [S^x,S^y] = i/2 $ leads to field operator
\begin{equation}
   \tilde \phi (i,t) = 
     S^x_i+iS^y_i = {1\over {\cal N}^{1/2}}
    \sum_{\vec k} a_{\vec k} e^{i\vec k \vec r_i-i\omega_{s}(\vec k)t}\;,\quad
\label{eq:letter.060}
\end{equation}
and the propagator of spin waves
\begin{equation}
    \chi(\xi,\xi') = -i \langle T \tilde\phi(\xi) \tilde\phi^\dagger(\xi')\rangle\;,\quad 
    \chi_{\omega, \vec k} = { 1 \over \omega - \omega_{s}(\vec k) + i\delta}
\label{eq:letter.070}
\end{equation}
Now we can put $z$ axis perpendicular to the plane of atomic lattice, as it was 
before spin wave quantization. The magnon dispersion for the Hamiltonian 
$J\vec S_i \vec S_{i'}$ is just $\omega_{s}(\vec k)=2J \vec \kappa_{\vec k}^2$, and we 
should preserve it since it should be consistent with interaction Hamiltonian.

\begin{figure}
\unitlength=1.00mm
\begin{picture}(76.00,60.00)
\put(-10.00,25.00){
\put(67.00,25){\makebox(0,0)[cb]{$D^{(0)}_{\omega,\vec k}$}}
\linethickness{0.4pt}
\put(61.00,20.00){\bf\oval(2.00,2.00)[t]}
\put(63.00,20.00){\oval(2.00,2.00)[b]}
\put(65.00,20.00){\oval(2.00,2.00)[t]}
\put(67.00,20.00){\oval(2.00,2.00)[b]}
\put(69.00,20.00){\oval(2.00,2.00)[t]}
\put(71.00,20.00){\oval(2.00,2.00)[b]}
\put(73.00,20.00){\oval(2.00,2.00)[t]}
\put(75.00,20.00){\oval(2.00,2.00)[b]}
\put(85.00,20.00){\makebox(0,0)[rc]{$+$}}
}
\put(-10.00,25.00){ }
\put(18.00,50){\makebox(0,0)[cb]{$D_{\omega,\vec k}$}}
\put(40.00,45.00){\makebox(0,0)[rc]{$=$}}
\linethickness{0.4pt}
\put(5.00,25.00){
\unitlength=1.00pt
\put(-0.4,-0.4){
\unitlength=1.00mm
\put(5.00,20.00){\oval(2.00,2.00)[t]}
\put(7.00,20.00){\oval(2.00,2.00)[b]}
\put(9.00,20.00){\oval(2.00,2.00)[t]}
\put(11.00,20.00){\oval(2.00,2.00)[b]}
\put(13.00,20.00){\oval(2.00,2.00)[t]}
\put(15.00,20.00){\oval(2.00,2.00)[b]}
\put(17.00,20.00){\oval(2.00,2.00)[t]}
\put(19.00,20.00){\oval(2.00,2.00)[b]}
}
\put(-0.2,-0.2){
\unitlength=1.00mm
\put(5.00,20.00){\oval(2.00,2.00)[t]}
\put(7.00,20.00){\oval(2.00,2.00)[b]}
\put(9.00,20.00){\oval(2.00,2.00)[t]}
\put(11.00,20.00){\oval(2.00,2.00)[b]}
\put(13.00,20.00){\oval(2.00,2.00)[t]}
\put(15.00,20.00){\oval(2.00,2.00)[b]}
\put(17.00,20.00){\oval(2.00,2.00)[t]}
\put(19.00,20.00){\oval(2.00,2.00)[b]}
}
\put(0.0,0.0){
\unitlength=1.00mm
\put(5.00,20.00){\oval(2.00,2.00)[t]}
\put(7.00,20.00){\oval(2.00,2.00)[b]}
\put(9.00,20.00){\oval(2.00,2.00)[t]}
\put(11.00,20.00){\oval(2.00,2.00)[b]}
\put(13.00,20.00){\oval(2.00,2.00)[t]}
\put(15.00,20.00){\oval(2.00,2.00)[b]}
\put(17.00,20.00){\oval(2.00,2.00)[t]}
\put(19.00,20.00){\oval(2.00,2.00)[b]}
}
\put(0.2,0.2){
\unitlength=1.00mm
\put(5.00,20.00){\oval(2.00,2.00)[t]}
\put(7.00,20.00){\oval(2.00,2.00)[b]}
\put(9.00,20.00){\oval(2.00,2.00)[t]}
\put(11.00,20.00){\oval(2.00,2.00)[b]}
\put(13.00,20.00){\oval(2.00,2.00)[t]}
\put(15.00,20.00){\oval(2.00,2.00)[b]}
\put(17.00,20.00){\oval(2.00,2.00)[t]}
\put(19.00,20.00){\oval(2.00,2.00)[b]}
}
\put(0.4,0.4){
\unitlength=1.00mm
\put(5.00,20.00){\oval(2.00,2.00)[t]}
\put(7.00,20.00){\oval(2.00,2.00)[b]}
\put(9.00,20.00){\oval(2.00,2.00)[t]}
\put(11.00,20.00){\oval(2.00,2.00)[b]}
\put(13.00,20.00){\oval(2.00,2.00)[t]}
\put(15.00,20.00){\oval(2.00,2.00)[b]}
\put(17.00,20.00){\oval(2.00,2.00)[t]}
\put(19.00,20.00){\oval(2.00,2.00)[b]}
}
}

\linethickness{0.4pt}
\bezier{200}(20.00,20.00)(40.00,35.00)(60.00,20.00)
\bezier{200}(20.00,20.00)(40.00,5.00)(60.00,20.00)
\put(38.39,27.43){\vector(1,0){3.05}}
\put(41.74,12.52){\vector(-1,0){3.40}}
\put(39.92,29.07){\makebox(0,0)[cb]{$\chi^\ast_{\omega - \omega',\vec q+\vec k/2}$}}
\put(39.98,9.99){\makebox(0,0)[ct]{$\chi_{\omega',\vec q -\vec k/2}$}}
\linethickness{0.4pt}
\put(61.00,20.00){\oval(2.00,2.00)[t]}
\put(63.00,20.00){\oval(2.00,2.00)[b]}
\put(65.00,20.00){\oval(2.00,2.00)[t]}
\put(67.00,20.00){\oval(2.00,2.00)[b]}
\put(69.00,20.00){\oval(2.00,2.00)[t]}
\put(71.00,20.00){\oval(2.00,2.00)[b]}
\put(73.00,20.00){\oval(2.00,2.00)[t]}
\put(75.00,20.00){\oval(2.00,2.00)[b]}
\linethickness{0.4pt}
\unitlength=1.00pt
\put(-0.4,-0.4){
\unitlength=1.00mm
\put(5.00,20.00){\oval(2.00,2.00)[t]}
\put(7.00,20.00){\oval(2.00,2.00)[b]}
\put(9.00,20.00){\oval(2.00,2.00)[t]}
\put(11.00,20.00){\oval(2.00,2.00)[b]}
\put(13.00,20.00){\oval(2.00,2.00)[t]}
\put(15.00,20.00){\oval(2.00,2.00)[b]}
\put(17.00,20.00){\oval(2.00,2.00)[t]}
\put(19.00,20.00){\oval(2.00,2.00)[b]}
}
\put(-0.2,-0.2){
\unitlength=1.00mm
\put(5.00,20.00){\oval(2.00,2.00)[t]}
\put(7.00,20.00){\oval(2.00,2.00)[b]}
\put(9.00,20.00){\oval(2.00,2.00)[t]}
\put(11.00,20.00){\oval(2.00,2.00)[b]}
\put(13.00,20.00){\oval(2.00,2.00)[t]}
\put(15.00,20.00){\oval(2.00,2.00)[b]}
\put(17.00,20.00){\oval(2.00,2.00)[t]}
\put(19.00,20.00){\oval(2.00,2.00)[b]}
}
\put(0.0,0.0){
\unitlength=1.00mm
\put(5.00,20.00){\oval(2.00,2.00)[t]}
\put(7.00,20.00){\oval(2.00,2.00)[b]}
\put(9.00,20.00){\oval(2.00,2.00)[t]}
\put(11.00,20.00){\oval(2.00,2.00)[b]}
\put(13.00,20.00){\oval(2.00,2.00)[t]}
\put(15.00,20.00){\oval(2.00,2.00)[b]}
\put(17.00,20.00){\oval(2.00,2.00)[t]}
\put(19.00,20.00){\oval(2.00,2.00)[b]}
}
\put(0.2,0.2){
\unitlength=1.00mm
\put(5.00,20.00){\oval(2.00,2.00)[t]}
\put(7.00,20.00){\oval(2.00,2.00)[b]}
\put(9.00,20.00){\oval(2.00,2.00)[t]}
\put(11.00,20.00){\oval(2.00,2.00)[b]}
\put(13.00,20.00){\oval(2.00,2.00)[t]}
\put(15.00,20.00){\oval(2.00,2.00)[b]}
\put(17.00,20.00){\oval(2.00,2.00)[t]}
\put(19.00,20.00){\oval(2.00,2.00)[b]}
}
\put(0.4,0.4){
\unitlength=1.00mm
\put(5.00,20.00){\oval(2.00,2.00)[t]}
\put(7.00,20.00){\oval(2.00,2.00)[b]}
\put(9.00,20.00){\oval(2.00,2.00)[t]}
\put(11.00,20.00){\oval(2.00,2.00)[b]}
\put(13.00,20.00){\oval(2.00,2.00)[t]}
\put(15.00,20.00){\oval(2.00,2.00)[b]}
\put(17.00,20.00){\oval(2.00,2.00)[t]}
\put(19.00,20.00){\oval(2.00,2.00)[b]}
}
\end{picture}

\caption{The Dyson equation for the collective phonon - two magnon excitation. 
         Vorteces have complicated $\vec k, \vec q$ dependence, roughly like $k^{2.5}$.}
\label{fig:diagram01}
\end{figure}

The interaction Hamiltonian is the sum over $\alpha = x,y$
\begin{equation}
       {\cal H} = \sum_{\alpha,\vec k,\vec q} g^\alpha_{\vec k, \vec q}
       b_{\vec k} a_{\vec q-\vec k/2} a^\dagger_{\vec q+\vec k/2 } 
       + \text{c.c.}
\label{eq:letter.080}
\end{equation}
where
\begin{equation}
       g^\alpha_{\vec k, \vec q} = 4g^\alpha_{\vec k}
        \kappa^\alpha_{\vec q-\vec k/2} \kappa^\alpha_{\vec q+\vec k/2}\;,\quad
      g^\alpha_{\vec k} = {dJ\over da} 
       { \kappa^\alpha_{\vec k}k^\alpha/k\over \sqrt{2 \rho \omega_0(\vec k)}}
\label{eq:letter.090}
\end{equation}
Then the Dyson equation for the phonon propagator becomes matrix,
\begin{eqnarray}
   \left[{1 \over  D_{\omega, \vec k}}\right]_{\alpha \beta} \!\!\!\! &=& 
   {1 \over  D^{(0)}_{\omega, \vec k}} \delta_{\alpha \beta} - M^{\alpha \beta}_{\omega \vec k}
\nonumber\\ 
   M^{\alpha \beta}_{\omega \vec k}  &=& 
     i \int { d\vec q d\omega' \over (2\pi)^3}
       g^\alpha_{\vec k, \vec q} g^\beta_{\vec k, \vec q}
       \chi_{\omega',\vec q -\vec k/2} \chi^\ast_{\omega - \omega',\vec q+\vec k/2} 
\label{eq:letter.100}
\end{eqnarray}
and integration of the magnon loop, see Fig.~\ref{fig:diagram01}, is our primary task.

The frequency integration is simple and keeping real part in the answer we have
\begin{equation}
    \int { d\omega' \over 2\pi i}   
    \chi_{\omega',\vec q -\vec k/2} 
    \chi^\ast_{\omega - \omega',\vec q+\vec k/2}    
    = { 1 \over \omega - 2J\vec \kappa_{2\vec q} \vec \kappa_{\vec k} }\;.
\label{eq:letter.110}
\end{equation}
The $\vec q$ integral in turn is complicated. At low frequencies important for phonon dispersion 
\begin{equation}
   M^{\alpha \beta}_{\omega \vec k}   
   = \omega  I^{\alpha\beta}_{\vec k} f_{\vec k}
   \;,\;\;
   I^{\alpha\beta}_{\vec k} = 
   g^\alpha_{\vec k } g^\beta_{\vec k } \,
   {\delta_{\alpha \beta}+\cos{k^\alpha a \over 2} \cos{k^\beta a \over 2} \over 2 \pi J^2 a^2 }\;;
\label{eq:letter.120}
\end{equation}
and it has the pole
\begin{equation}
   f_{\vec k} = 
    {   \sin^2 {k^y a\over 2} / \sin^2 {k^x a\over 2}
      \over \sin^2 {k^x a\over 2}  -  \sin^2 {k^y a\over 2}} 
    + \text {non-singular term} \;, 
\label{eq:letter.130}
\end{equation}
where $\sin^2 {k^x a\over 2} > \sin^2 {k^y a\over 2}$ and one should interchange $k^x $ with $ k^y$ when
$\sin^2 {k^x a\over 2} < \sin^2 {k^y a\over 2}$.

There are two branches of the phonon dispersion, as obtained from 
Eqs.~(\ref{eq:letter.100}), (\ref{eq:letter.120}), and 
$\text{det} D^{-1}_{\omega, \vec k} = 0$. Near the pole of $f_{\vec k}$ 
one branch is the original phonon dispersion $\omega=\omega_0(\vec k)$ and another branch is 
\begin{equation}
   {\omega^2-\omega^2_0(\vec k) \over ( \vec \kappa_{\vec k} )^2 } - \omega 
   [I^{xx}_{\vec k} + I^{yy}_{\vec k}]  f_{\vec k} + \text{n.s.t.} = 0
\label{eq:letter.140}
\end{equation}
where n.s.t. stands for non-singular terms. Our result Eq.~(\ref{eq:intro.010}) with all
prefactors reads
\begin{equation}
   \omega = (\vec \kappa_{\vec k} )^2 [I^{xx}_{\vec k} + I^{yy}_{\vec k}]  f_{\vec k}
\label{eq:letter.150}
\end{equation}
valid for $\omega \ll J (\vec \kappa_{\vec k} )^2$ - condition for $\omega$-expansion of 
$M^{\alpha\beta}_{\omega \vec k}$.

In conclusions the phonon dispersion is modified by interaction with spin waves; the effect is strong for large 
phonon wave numbers (new branch goes $\propto k^3$) and should be observed experimentally. The phonon dispersion
becomes singular along $k_x = \pm k_y$ lines opening a lot of room for electron pairing, that could be key 
solution for high-temperature superconductivity. This work has been inspired by discussion at Physics department of
Stanford University. The experiment showed that low $k$ part of the phonon spectrum depends strongly on electron density 
where as high $k$ part is completely insensitive. This is in agreement with our results, because the
renormalization of $D^{(0)}_{\omega \vec k}$ by electron-phonon interaction does not change $\omega(\vec k)$ given
by Eq.~(\ref{eq:letter.150}).




\end{document}